\documentclass[12pt,a4]{article}
\usepackage{mcite}
\usepackage{epsfig}
\usepackage{axodraw}

\epsfverbosetrue

\def\3{\ss}

\newcommand{\tev}{{\rm Te}\kern-1.pt{\rm V}}
\newcommand{\gev}{{\rm Ge}\kern-1.pt{\rm V}}
\newcommand{\mev}{{\rm Me}\kern-1.pt{\rm V}}
\newcommand{\kev}{{\rm Ke}\kern-1.pt{\rm V}}
\newcommand{\gevsq}{\mbox{$\mathrm{{\rm Ge}\kern-1.pt{\rm V}}^2$}}
\newcommand{\gevmsq}{\mbox{$\mathrm{{\rm Ge}\kern-1.pt{\rm V}}^{-2}$}}

\newcommand{\mayor} {\mbox{\raisebox{-0.4ex}
{$\;\stackrel{>}{\scriptstyle \sim}\;$}}}

\begin{document}
\begin{titlepage}

\begin{center}
\begin{huge}
\bf Prospects for Pentaquark Searches in $e^+e^-$ Annihilations and $\gamma\gamma$ Collisions \\
\end{huge}

\vspace{2.cm}

\Large S.~Armstrong, B.~Mellado, Sau Lan Wu \\
\vspace{0.5cm}
{\Large\it Department of Physics \\
University of Wisconsin - Madison \\
   Madison, Wisconsin 53706 USA }

\vspace{1.5cm}

\begin{abstract}
\noindent Recent strong experimental evidence of a narrow exotic S
= +1 baryon resonance, $\Theta^+$, suggests the existence of other
exotic baryons. We discuss the prospects of confirming earlier
experimental evidence of $\Theta^+$ and the observation of
additional hypothetical exotic baryons in $e^+e^-$ annihilations
and $\gamma\gamma$ collisions at LEP and B~Factories.
\end{abstract}
\end{center}
\setcounter{page}{0}
\thispagestyle{empty}

\end{titlepage}

\newpage

\pagenumbering{arabic}

\section{Introduction}
\label{sec:introduction}

Recent strong experimental evidence  of a narrow exotic baryon
resonance with strangeness $S=+1$, $\Theta^+$, may open a new
chapter in the development of hadron spectroscopy. A narrow
resonance peak near $1540\,\mev/c^2$ was observed in $pK^0_s$ and
$nK^+$ invariant mass spectra in a number of different
reactions~\cite{pan_66_1715,prl_91_012002,hep-ex_0307018,hep-ex_0311046,hep-ex_0307083,hep-ex_0309042,hep-ex_0312044}.
However, it has been argued recently that the excess of events
observed in the $K^+n$ invariant mass spectrum may be caused by
kinematic reflections from the decay of mesons such as
$f_2(1275)$, $a_2(1320)$, and
$\rho_3(1690)$~\cite{hep-ph_0311125}.

The quark model based on quark-antiquark mesons and three-quark
baryons has been greatly successful in understanding a wealth of
hadron spectroscopy. However, there is nothing in Quantum
Chromodynamics (QCD) that prevents pentaquark bound states or
resonances to appear in Nature. If fully confirmed, this exotic
baryon would be most naturally explained as a pentaquark state of
four quarks and an antiquark $(udud\overline{s})$.

According to the Standard Model (SM), baryons are arranged in
octects with spin-${1 \over 2}$ and decuplets with spin-${3 \over
2}$ . In the mid 1980s, the existence of anti-decuplet was
hypothesized within the framework of the chiral soliton
model~\cite{Diak_Petrov_1984,np_256_600,Prasza_1987,Walliser_1992,np_A548_649}.
In 1987, Praszalowicz presented the first estimate of the mass of
what is today $\Theta^+$ (viewed as the lightest member of the
hypothesized anti-decuplet), a value around $1530\,\mev/c^2$, in
striking agreement with the
measurement~\cite{Prasza_1987_talk,hep-ph_0308114}. In 1997,
Diakonov, Petrov and Polyakov calculated the mass and the width of
the lightest member of the anti-decuplet operating under the
assumption that the known nucleon resonance $N(1710,{1 \over
2}^+)$ is a member of the hypothesized
antidecuplet~\cite{zp_A359_305}.\footnote{The authors used
initially the notation $Z^+$. The later was renamed by the authors
to $\Theta$~\cite{Diakonov_1}.} They predicted $\Theta^+$ to have
a mass around $1530\,\mev/c^2$ and a width less than
$15\,\mev/c^2$. Predictions were made for the mass, width and the
branching ratios of other members of the suggested anti-decuplet
(i.e., $N$, $\Sigma$,  and $\Xi$
baryons)~\cite{zp_A359_305,hep-ph_0310212}.

No accurate prediction of the $\Theta^+$ mass and width was ever
made within the constituent quark model. The direct determination
of the $\Theta^+$ mass jointly with upper limits on the decay
width may be used as input to models based on the constituent
quark model. An incresing number of models are apprearing, which
consider the mass, width, spin, isospin and parity of $\Theta^+$
and make predictions concerning the existence of additional
baryons (for a recent collection of references
see~\cite{hep-ph_0312325}). The mass and the quantum numbers of
the $\Theta^+$ may be calculated using lattice QCD, which is known
to reproduce mass ratios of stable hadrons. Quenched lattice QCD
computations of $\Theta^+$ mass agree very well with the
experimental value~\cite{jhep_0311_70,hep-lat_0310014}.


Recently, the NA49 Collaboration reported  evidence (roughly 4
standard deviations) of a narrow resonance in the $\Xi^-\pi^-$
invariant mass spectrum observed in proton-proton collisions at
$\sqrt{s} = 17.2\,\gev$~\cite{hep-ex_0310014}. A less prominent
excess of events were reported in the $\Xi^-\pi^+$,
$\overline{\Xi}^+\pi^-$ and $\overline{\Xi}^+\pi^+$ invariant mass
spectra. These hint, nevertheless, to the existence of a isospin
quartet, as predicted by the anti-decuplet idea. Jaffe and Wilczek
have interpreted the results of the NA49 Collaboration as an
evidence of an octet cascade nearly degenerate with the resonance
observed in the  $\Xi^-\pi^-$ system~\cite{hep-ph_0312369}.

The existence of heavier exotic baryons, where the anti-strange
quark, $\overline{s}$, is replaced by anti-charm or anti-beauty,
now seems to be a natural consequence of the observation of
$\Theta^+$. The masses of the lightest members of higher exotic
multiples ($\Theta^0_c$ and $\Theta^+_b$) have been recently
conjectured by a number of authors based on the constituent quark
model~\cite{hep-ph_0307341,hep-ph_0307343,hep-ph_0308176,hep-ph_0312319}.
These models suggest that $\Theta^0_c$ and $\Theta^+_b$ will
appear as relatively narrow resonances. This is in contradiction
with a recent calculation performed with quenched lattice QCD,
which predicts that the $\Theta^0_c$ mass is $640\,\mev/c^2$ above
the DN (D-meson and N-baryon) threshold~\cite{hep-lat_0310014}.

In the present discussion, we present the prospects of confirming
earlier experimental evidence of $\Theta^+$ and the observation of
additional hypothetical exotic baryons in $e^+e^+$ collisions
provided by LEP and B~Factories. The potential of observing
exotic baryons in $\gamma\gamma$ collisions and $e^+e^-$
annihilations is discussed.

\section{Pentaquark Production}
\label{sec:production}

Little is known about the production mechanism giving rise to
exotic baryon states in collisions involving hadrons. One can make
educated guesses given the fact that $\Theta^+$ has been observed
in a number of different reactions.

\begin{figure}[t]
{\centerline{\epsfig{figure=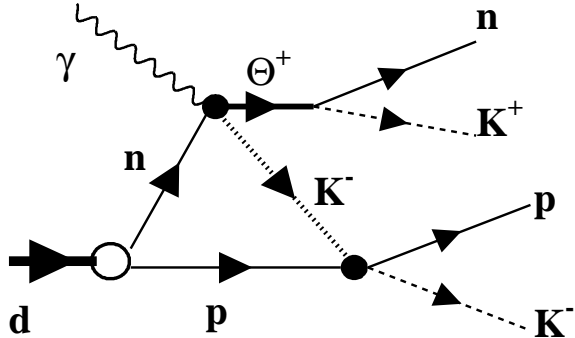,width=8cm}}}
\caption[]{A re-scattering diagram in $\gamma d$ collisions that
could contribute to the exclusive reaction mechanism leading to
$\Theta^+\rightarrow K^+n$ and an energetic proton. The exotic
baryon is produced via interactions in the final state (after the
CLAS Collaboration).} \label{fig:gn}
\end{figure}

Collisions involving photons are attractive for
$\Theta^+\rightarrow KN$ searches thanks to the relatively high
content of strange quarks in photons. Most of the reactions in
which $\Theta^+$ baryons have been observed involve photon collisions with
hadrons.

At low energies in the exclusive reaction $\gamma d\rightarrow
K^+K^-pn$ where $\Theta^+\rightarrow K^+n$, one could draw a
re-scattering diagram, as depicted in Figure~\ref{fig:gn}. Thanks
to the re-scattering process in the final state, the proton, which
acts as an spectator, acquires enough momentum that it may be
detected~\cite{hep-ex_0307018}.

\begin{figure}[t]
{\centerline{\epsfig{figure=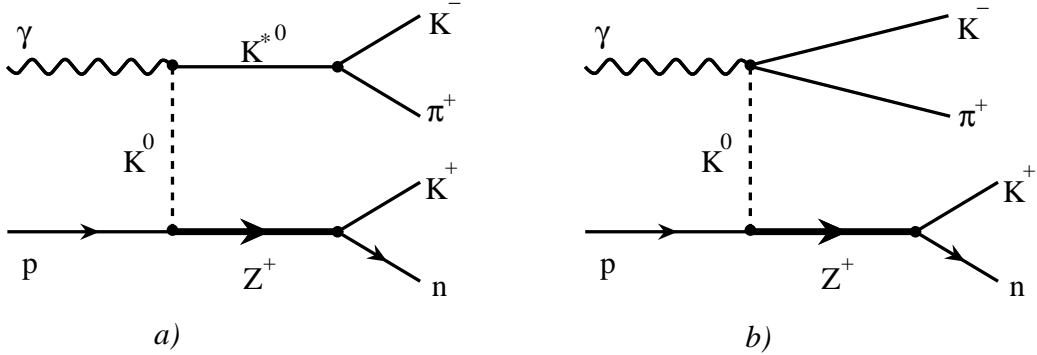,width=18cm}}}
\caption[]{Feynman diagrams for $\gamma p\rightarrow \pi^+K^+K^-n$
(after V.~Kubarovsky and S.~Stepanian, for the CLAS
Collaboration).} \label{fig:gp}
\end{figure}

Figure~\ref{fig:gp} depicts possible Feynman diagrams for the
reaction $\gamma p\rightarrow\pi^+K^+K^-n$. These diagrams
correspond to a $t$-channel exchange mechanism. Hence, one would
expect that the  $\cos{\theta^*}$ distribution, where $\theta^*$
is the angle between the $\pi^+K^-$ momentum and the photon beam
in the center of mass system, peaks in the forward region. This
feature has been verified experimentally~\cite{hep-ex_0307088}.

\begin{figure}[t]
{\centerline{\epsfig{figure=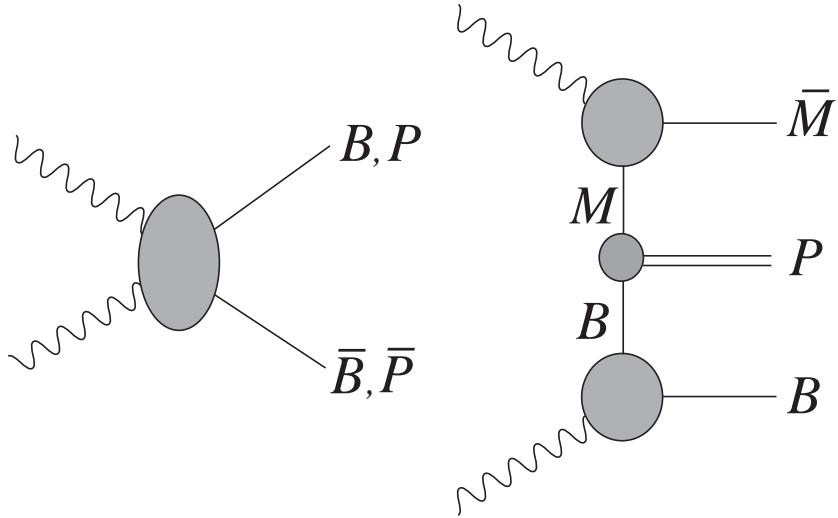,width=11cm}}}
\caption[]{The diagrams on the left and right correspond to single
and double $\Theta^+$ production in a $\gamma\gamma$ collision.
The blobs imply non-perturbative probability amplitudes. $B
(\overline{B})$, $M (\overline{M})$ and $P$ ($\overline{P}$)
denote baryon, meson and pentaquark, respectively.} \label{fig:gg}
\end{figure}

Figure~\ref{fig:gg} depicts possible diagrams for the production of
pentaquarks in $\gamma\gamma$ collisions. The diagrams on the left
and right in Figure~\ref{fig:gg} correspond to the exclusive
double and single pentaquark production. Blobs are depicted in
both diagrams; these imply the presence of non-perturbative
probability amplitudes.

Cross sections for the exclusive production
$\gamma\gamma\rightarrow p\overline{p}$,
$\Sigma^0\overline{\Sigma}^0$, $\Lambda^0\overline{\Lambda}^0$
collisions have been reported by a number of experiments:
CLEO~\cite{pr_50_5484}, VENUS~\cite{pl_407_185},
OPAL~\cite{epj_28_45}, L3~\cite{pl_536_24,pl_571_11}, and
BELLE~\cite{Kuo_photon2003}. The agreement between the various
experiments is not excellent. Nevertheless, the cross section of
$\gamma\gamma\rightarrow p\overline{p}$ for $\gamma-\gamma$
center-of-mass energy, $W_{\gamma\gamma}\approx 2 M_{p}$, where
$M_p$ corresponds to the proton mass, is of the order of 4 to
8$\,$nb.\footnote{Cross sections of $\gamma\gamma\rightarrow
\Lambda^0\overline{\Lambda}^0$ at $W_{\gamma\gamma}\mayor 2
M_{\Lambda^0}$ reported by CLEO and L3 differ by an order of
magnitude.} Thanks to the high luminosity regime, experiments at
B~Factories are in an excellent position to collect large
statistics of di-baryon events in exclusive $\gamma\gamma$
reactions.\footnote{Belle has collected approximately $2\,$x$10^4
~p\overline{p}$ candidates and roughly 6x$10^{3}$
$\Sigma^0\overline{\Sigma}^0$, $\Lambda^0\overline{\Lambda}^0$
pairs with approximately $89\,$fb$^{-1}$ of integrated
luminosity~\cite{Kuo_photon2003}.} Given the cross sections for
di-baryon production being of the order of few nb, di-pentaquark
production could be observed at the B~Factories provided that the
reaction $\gamma\gamma\rightarrow \Theta^+\overline{\Theta}^-$ is
suppressed with respect to $\gamma\gamma\rightarrow p\overline{p}$
by less than three orders of magnitude.

The exclusive production of di-baryons in $\gamma\gamma$ collision
may be viewed within the framework of the hard scattering picture
(HSP), according to which the scattering amplitude may be
expressed in terms of a convolution of a process-dependent
(calculable in perturbation theory) piece with a
process-independent amplitude for finding the corresponding hadron
in the final state~\cite{pr_22_2157,Lepage_Brodsky_1989}.
Experimental results seem to be reasonably well described by this
approach~\cite{epj_28_249,hep-ph_0309095}. These models operate
under the assumption that baryons can be expressed in terms of
quark structures (quark-di-quark bound states). A number of groups
are able to interpret the pentaquarks in terms of bound states of
quark structures: Jaffe and Wilczek view the pentaquark as a bound
state of diquark-diquark-antiquark~\cite{hep-ph_0307341}; Karliner
and Lipkin advocate a diquark-triquark
configuration~\cite{hep-ph_0307343}. In principle, the formalism
developed within the HSP may be extended to accommodate exclusive
di-pentaquark production in $\gamma\gamma$
collisions~\cite{Schweiger_1}.

Single pentaquark production may be viewed as a collision of
non-resonant~\footnote{By non-resonant it is implied that the
baryons and mesons are not produced as a result of the decay of a
heavier baryons.} di-baryon and a di-meson pairs (see
Figure~\ref{fig:gg}). One of the incoming photons fluctuates into
two mesons with strangeness. One of them collides with the
hadronic system produced by the other incoming photon.
Non-resonant exclusive production of $p\overline{p}\pi^+\pi^-$ was
observed by the TPC/Two-Gamma Collaboration in $\gamma\gamma$
collisions at
PEP~\cite{pr_40_2772}.\footnote{In~\cite{pr_40_2772} it was
reported that no $\Delta^0\overline{\Delta}^0$ production was
observed and only a small fraction of the events were consistent
with $\gamma\gamma\rightarrow\Delta^{++}\overline{\Delta}^{--}$,
$\Delta^{++}\overline{p}\pi^-$, or
$\overline{\Delta}^{--}p\pi^+$.} The $W_{\gamma\gamma}$ dependence
of the cross section was parameterized with a power-law functional
form including a threshold effect~\cite{pr_40_2772}
\begin{equation}
\sigma\left( \gamma\gamma\rightarrow p\overline{p}\pi^+\pi^-
\right) \propto W_{\gamma\gamma}^{-b}\left[ 1-4{\left(
M_p+M_{\pi}\right)^2 \over W_{\gamma\gamma}^2 }\right]^{1 \over 2},
\label{eq:pppp}
\end{equation}
where $M_{\pi}$ is the mass of the $\pi$ meson and $b$ is a free
parameter. A fit to the data yielded $b\approx 5$.

It is relevant to note that no significant
$\rho^0\rightarrow\pi^+\pi^-$ contribution was found in
$p\overline{p}\pi^+\pi^-$ events. We will assume that the
contribution from $p\overline{p}\phi$ in $p\overline{p}KK$ events
is also negligible. Furthermore, one can assume that the exclusive
non-resonant cross sections $\sigma\left( \gamma\gamma\rightarrow
p\overline{p}K^0\overline{K}^0\right) \approx \sigma\left( \gamma\gamma\rightarrow
p\overline{p}K^+K^-\right)$. By
using factorization of vertexes in the right diagram  in
Figure~\ref{fig:gg} and Expression~\ref{eq:pppp}, one can verify:
\begin{equation}
\sigma\left( \gamma\gamma\rightarrow p\overline{p}K^+K^-\right)
\approx \sigma\left( \gamma\gamma\rightarrow
p\overline{p}\pi^+\pi^- \right)  {\sigma\left(
\gamma\gamma\rightarrow \pi^+\pi^-K^+K^-\right)  \over
\sigma\left( \gamma\gamma\rightarrow \pi^+\pi^-\pi^+\pi^- \right)}
\left( {M_p+M_{\pi} \over M_p+M_{K} }\right)^5,
\label{eq:ppkk}
\end{equation}
where $\sigma\left( \gamma\gamma\rightarrow
p\overline{p}K^+K^-\right)$ and $\sigma\left(
\gamma\gamma\rightarrow p\overline{p}\pi^+\pi^- \right)$ are
evaluated for $W_{\gamma\gamma}$ close the corresponding
production threshold. Here, $M_K$ is the mass of the $K$ meson.

Results of four-prong final states (including
$\pi^+\pi^-\pi^+\pi^-$ and $\pi^+\pi^-K^+K^-$) in $\gamma\gamma$
were observed in
collisions~\cite{pr_37_28,zp_C41_353,zp_C38_521,pl_198_255,zp_C32_11}.
The $\gamma\gamma\rightarrow \pi^+\pi^-\pi^+\pi^-$ is dominated by
the $\gamma\gamma\rightarrow\rho^0\rho^0$ for $W_{\gamma\gamma}$
close to the production threshold but it is negligible for
$W_{\gamma\gamma}>2\,\gev$. The contribution to
$\gamma\gamma\rightarrow \pi^+\pi^-K^+K^-$ from
$\gamma\gamma\rightarrow \pi^+\pi^-K^+K^-$ from
$\gamma\gamma\rightarrow\pi^+\pi^-\phi$ is very small as it is
concentrated at the production threshold. The amplitude squared of
the matrix elements in the reactions $\gamma\gamma\rightarrow
\pi^+\pi^-\pi^+\pi^-$ and $\gamma\gamma\rightarrow
\pi^+\pi^-K^+K^-$ were extracted from the cross sections by
dividing by the phase space and the photon flux factors. The
energy dependence of both matrix elements is similar for
$W_{\gamma\gamma}>2\,\gev$, suggesting similar production
mechanisms. The ratio ${\sigma\left( \gamma\gamma\rightarrow
\pi^+\pi^-K^+K^-\right)  \over \sigma\left(
\gamma\gamma\rightarrow \pi^+\pi^-\pi^+\pi^-\right)}\approx
0.25$~to~$0.3$ for
$W_{\gamma\gamma}>2\,\gev$~\cite{pr_37_28,zp_C41_353,zp_C38_521,pl_198_255,zp_C32_11}.
Using Expression~\ref{eq:ppkk} we obtain $\sigma\left(
\gamma\gamma\rightarrow p\overline{p}K^+K^-\right)\approx 0.1\,$nb
for $W_{\gamma\gamma}\mayor2\left(M_p+M_K\right)$. The experiments
at B~Factories could be able to reconstruct on the order of
hundreds of $p\overline{p}$-$K^0\overline{K}^0$ pairs in
$\gamma\gamma$ collisions.

One might believe that exotic baryons, such as $\Theta^+$ and
other members of the hypothetical anti-decuplet, may only be
observed in reactions involving $\gamma$'s and hadrons
($\gamma\gamma$, $NN$, $\gamma N$, $\pi N$, $KN$ ). However, a
narrow resonance consistent with $\Theta^+$ has been also observed
in the $K^0_sp$ spectrum in charge current neutrino experiments
($\nu_{\mu} CC$ and
$\overline{\nu}_{\mu}CC$)~\cite{hep-ex_0309042}. In this reaction,
$\Theta^+$ is most probably produced in the hadronization process.
According to~\cite{hep-ex_0309042}, the appearance of most of
$\Theta^+$ candidates occurred in the reaction
$\nu_{\mu}Ne\rightarrow K^0_spX$. Very few candidates were
observed in the similar reaction on Hydrogen and Deuterium,
despite the fact that the number of charged-current interactions
used in their analysis was a factor of 2.5 larger than that
available for charged-current interaction with Neon.

The observation of $\Theta^+\rightarrow K^+p$ in neutrino-induced
charged-current interactions gives hope that exotic baryons may be
observed in hadronic $Z$ decays with LEP data at $e^+e^-$
center-of-mass energies  near $91\,\gev$. Generally speaking,
inclusive pentaquak searches may be performed on
$e^+e^-\rightarrow q\overline{q}$ at LEP and B~Factories.
Additionally, it has been recently argued that $\Theta^+$ and
exotic charmed baryons may be observed in $B$ meson
decays~\cite{hep-ph_0312269}.

Together with the observation of $\Theta^+\rightarrow KN$, three
experiments have reported the production of
$\Lambda^0(1520)\rightarrow
KN$~\cite{hep-ex_0307018,hep-ex_0307083,hep-ex_0312044}. Similar
production rates have been reported for $\Theta^+$ and
$\Lambda^0(1520)$. This is understandable due the presence of
strangeness and their proximity in
mass.\footnote{$\Theta^+(1540)$ displays $S=+1, I(J^P)=0({1 \over
2}^+)$ (the spin and parity have not been established
experimentally yet), whereas $\Lambda^0(1520)$ complies with
$S=-1, I(J^P)=0({3 \over 2}^+)$ with a full width of
$15.6\,\mev/c^2$~\cite{prl_66_010001}.} Unfortunately, the
production rate of $\Lambda^0(1520)\rightarrow K^-p$ was not
reported in the neutrino-induced charged-current
reactions~\cite{hep-ex_0309042}.

Pair or single production of heavier pentaquarks, such as
$\Theta^0_c$ and $\Theta^+_b$, may be difficult to produce in
$\gamma\gamma$ collisions due to the relatively small content of
heavy quarks in the photon. On the contrary, inclusive searches
may be performed because continuum $e^+e^-$ annihilations and
hadronic $Z$ decays include decays into heavy quarks (i.e.,
$c\overline{c}$, $b\overline{b}$) (see
Section~\ref{sec:prospective}).

Exclusive production of pentaquarks in $e^+e^+$ annihilations is
expected to drop very rapidly with the center-of-mass
energy~\cite{hep-ph_0311355}. In order to observe exclusive
pentaquark production in $e^+e^+$ annihilations, the
center-of-mass energy should be somewhat above twice the
pentaquark mass (see Section~\ref{sec:prospective}).

\section{$e^+e^-$ Collisions at LEP and B~Factories}
\label{sec:LEPBFAC}

From 1989 to 1995, LEP accelerated
$e^+$ and $e^-$ beams to an energy of roughly $45\,\gev$, allowing
collisions with center-of-mass energies at or near the $Z$ peak.
This first phase of LEP (LEP1) delivered a total of 200 pb$^{-1}$
to each detector experiment (ALEPH, DELPHI, L3, and OPAL). In
these data, each experiment recorded more than 3 million hadronic
$Z$ decays yielding in excess of 5x10$^5$ reconstructed $K^0_S$
mesons and $10^4$ $\Lambda^0$ baryons.

The second phase of LEP (LEP2) involved increasing the
center-of-mass energy to nearly $210\,\gev$ in consecutive annual
upgrades between 1995 and 2000. In November of 1995, LEP attained
center-of-mass energies between 130 and $136\,\gev$. Subsequent
years yielded energies of 161 and $172\,\gev$ (1996), $183\,\gev$
(1997), $189\,\gev$ (1998), between 192 and $196\,\gev$ (1999),
and between 200 and $209\,\gev$ (2000). An integrated luminosity
of about 500\,pb$^{-1}$ was delivered during LEP2 to each
experiment.

The CESR accelerator at Cornell collided $e^+e^-$ with a
center-of-mass energy at or just below the $\Upsilon(4S)$ mass
$(10.58\,\gev)$. By 1999 the CLEOII~\cite{nim_320_66} detector had
collected over $9.1\,$fb$^{-1}$ on resonance and $4.4\,$fb$^{-1}$
just below the resonance.\footnote{In 1995 CLEOII was upgraded to
CLEOII.V.} During 2001-2003 the CESR accelerator delivered to
CLEOIII~\cite{nim_384_61} several fb$^{-1}$ at center-of-mass
close to all $\Upsilon$ resonances. In 2003 the CESR-c program
started by running at $\psi(3770)$~\cite{CLNS_01_1742}.

The PEP-II accelerator at SLAC is a high-luminosity asymmetric
$e^+e^-$ collider with a center-of-mass energy near that
$\Upsilon(4S)$ peak with electron and positron beam energies of
$9\,\gev$ and $3.1\,\gev$, respectively.  Thus far, data-taking of
the BaBar experiment~\cite{BaBarTech} consists of three runs:
91.6\,fb$^{-1}$ of data were collected in Runs 1 and 2; a further
33.8\,fb$^{-1}$ were collected in Run~3.

Like PEP-II, KEKB is a dedicated high-luminosity asymmetric
$e^+e^-$ collider with a center-of-mass energy near the
$\Upsilon(4S)$ peak with electron and positron beam energies of
$8\,\gev$ and $3.5\,\gev$, respectively~\cite{KEKB}.  Between
October 1999 and December 2003, the Belle~\cite{nim_479_117}
experiment at KEKB received an integrated luminosity of
177.2\,fb$^{-1}$.

LEP and B~Factories offer the possibility of investigating
$\gamma\gamma$ collisions in the process $e^+e^-\rightarrow
e^+e^-X$, where the flux factor of $\gamma$ radiated of the
incoming $e^+e^-$ are well known
functions~\cite{jetp_14_1308,pr_15_181}. The outgoing leptons
carry most of the energy of the incoming beams and their
transverse momenta are usually very small, thus they scape through
the beam-pipe undetected. Hence, the photon virtuality is small
enough so that the colliding photons in untagged events may be
treated as quasi-real.

Due to baryon number conservation, baryons need to be pair
produced in a $\gamma\gamma$ collision. We will be particularly
interested in observing the exclusive reaction $e^+e^-\rightarrow
e^+e^-B\overline{B}M\overline{M}$ (see
Section~\ref{sec:production}). In order to observe $\Theta^+$ and
$\Xi_{3 \over 2}$ exotic baryons,
$W_{\gamma\gamma}\mayor3$~to~$4\,\gev$, which is kinematically
accessible at LEP and B~Factories.

The potential of the B factories is also discussed
in~\cite{hep-ph_0312269}.

\section{Prospective Searches}
\label{sec:prospective}

In this Section, we compile a list of final states that need to be
looked into with data collected at LEP and B~Factories. The
scope of this exercise is
\begin{itemize}
\item Confirming the discovery of $\Theta^+$ reported by other experiments. Make a measurement of the mass and width of this exotic baryon;
\item Searching for heavier partners of $\Theta^+$, in which $\overline{s}$ is replaced by $\overline{c}$ and $\overline{b}$ ($\Theta^+_c$ and $\Theta^+_b$, respectively);
\item Searching for a isospin quartet $\Xi_{3 \over 2}$ members of the hypothetical anti-decuplet, which existence has been hinted by NA49~\cite{hep-ex_0310014};
\item Pursuing a more general, model-independent, search for resonances in baryon-meson invariant mass spectra.
\end{itemize}

There are no theoretical predictions of the production rate of
exotic baryons in $e^+e^-$ annihilations and $\gamma\gamma$
collisions. However, the observation of $\Theta^+$ may
serve as a "calibration" process: exotic baryons with
strangeness may be produced copiously enough in $e^+e^-$
annihilations and $\gamma\gamma$ collisions at LEP and the
B~Factories. The production rate of the known baryon resonance
$\Lambda^0(1520)\rightarrow KN,\Sigma\pi$ may also give a feeling
for the expected production rate of $\Theta^+$.

\begin{table}[t]
\begin{center}
\begin{tabular}{||c|c||c|c||}
\hline Channel   & Mass Range $(\mev/c^2)$ & $e^+e^-\rightarrow q\overline{q}$
& $\gamma\gamma$ \\
\hline
\hline $\Theta^+\rightarrow K_s^0p \rightarrow p\pi^+\pi^-$ & 1530-1540 &  OK  &  OK    \\
\hline $\Theta^{*+}\rightarrow K_s^0p \rightarrow p\pi^+\pi^-$ & 1540-1680 &  OK  &  OK    \\
\hline
\hline $\Theta^0_c\rightarrow D^-p$ & 2700-3000 &  OK  &  ?    \\
\hline $\Theta^0_c\rightarrow \Theta^+\pi^-\rightarrow p\pi^+\pi^-\pi^-$ & 2700-3000 &  OK  &  ?    \\
\hline $\Theta^+_b\rightarrow B^0p$ & 6000-6400 &  OK  &  ?    \\
\hline $\Theta^+_b\rightarrow\Theta^0_c\pi^+$ & 6000-6400 &  OK  &  ?    \\
\hline \hline $\Xi_{3 \over 2}\rightarrow \Xi(\rightarrow
\Lambda^0\pi)\pi\rightarrow p\pi\pi\pi$ &
$\approx 1860$ & OK & OK  \\
\hline $\Xi^{\pm}_{3 \over 2}\rightarrow \Xi^0(\rightarrow
\Lambda^0\pi^0) \pi^\pm \rightarrow p\pi^-\pi^0\pi^\pm$ & $\approx 1860$ & OK & OK \\
\hline $\Xi^{\pm}_{3 \over 2}\rightarrow \Sigma^0(\rightarrow
p\pi^0) K^\pm \rightarrow p\pi^-\pi^0K^\pm$ & $\approx 1860$ & OK & OK \\
\hline $\Xi^{+}_{3 \over 2}\rightarrow \Sigma^+(\rightarrow
p\pi^0) K_S^0 \rightarrow p\pi^0\pi^+\pi^-$ & $\approx 1860$ & OK & OK \\
\hline $\Xi^{0}_{3 \over 2}\rightarrow \Sigma^+ K^- \rightarrow
p\pi^0 K^-$  & $\approx 1860$ & OK & OK  \\
\hline $\overline{\Xi}^{++}_{3 \over 2}\rightarrow \Sigma^+ K^+ \rightarrow
p\pi^0 K^+$  & $\approx 1860$ & OK & OK  \\
 \hline
 \end{tabular}
\caption{List of prospective channels for exotic Baryon searches.
It is assumed that the LEP and B~Factories experiments cannot reconstruct
neutrons with good efficiency.}
 \label{tab:channels}
\end{center}
\end{table}

Table~\ref{tab:channels} displays a list of prospective final
states for exotic baryon searches. It is assumed that the
experiments at LEP and B~Factories are incapable of reconstructing
neutrons with good efficiency; hence these channels are excluded
from the discussion.

The first row in Table~\ref{tab:channels} implies that the known
exotic baryon $\Theta^+$ may be searched for in both $e^+e^-$
annihilations and $\gamma\gamma$ collisions. The second row in
Table~\ref{tab:channels} corresponds to an orbital excitation of
$\Theta^+$, $\Theta^{*+}$ with $J^P={3 \over 2}^+$ (as opposed to
${3 \over 2}^+$ for $\Theta^+$) predicted
by~\cite{hep-ph_0311258}. The third and fourth rows in Table 1
correspond to lightest members of hypothetical higher multiplets.

Rows 3 through 6 correspond to heavy pentaquark decay modes. The
mass predictions given
in~\cite{hep-ph_0307341,hep-ph_0307343,hep-ph_0308176} are used.
Rows 4 and 6 correspond to heavy pentaquark weak decays, as
conjectured in~\cite{hep-ph_0312319}.

B~Factories are also charm factories, since $c\overline{c}$ pairs
are produced as copiously as $b\overline{b}$. Large statistics of
$D$ mesons have been reconstructed at
B~Factories~\cite{prl_89_251802,hep-ex_0308034,prl_91_121801,pr_65_091104},
turning them into attractive scenarios for $\Theta^0_c$ searches.
As pointed out in~\cite{hep-ph_0312269}, charmed pentaquarks may
be produced in B decays. Additionally, the branching ratio of
$B^0\rightarrow D^*(2010)^\pm p\overline{p}\pi^\mp$ is
6.5x10$^{-4}$~\cite{prl_66_010001}, allowing the final state
$D^*(2010)^\pm p\overline{p}$ to be produced in B decays copiously
enough at B~Factories. Unfortunately, because of the limited
center-of-mass energies provided by B~Factories, the production of
$\Theta^+_b$ there is not kinematically accessible. LEP
experiments are suited for heavy exotic baryon searches in
$e^+e^-$ annihilations.\footnote{For instance, the ALEPH
Collaboration reported on the observation of excited charm and
beauty states~\cite{zp_C69_393,zp_C73_601,pl_526_34}. These
analyses may be trivially turned into pentaquark searches.} A
question mark is placed under $\gamma\gamma$ collisions, since the
$c\overline{c}$ and $b\overline{b}$ content of the photons is
significantly smaller than that of lighter quarks.

The following rows in Table~\ref{tab:channels} correspond to
prospective searches of the isospin quartet $\Xi_{3 \over 2}$. In
the seventh row the charge of  $\Xi_{3 \over 2}$ is not specified.
Here, $\Xi_{3 \over 2}^{--}$, $\Xi_{3 \over 2}^0$ and their
antiparticles are implied. It is worth noting that the CESR
accelerator has recently started delivering $e^+e^-$ collisions at
a center-of-mass close to the mass of $\psi(3770)$, which is just
above the threshold for double $\Xi_{3 \over 2}$ production.

The prospective final states given in Table~\ref{tab:channels} are
motivated by theoretical expectations, which are more or less
model-dependent. From an experimentalist's point of view, a more
general search for baryon-meson resonances may be pursued. We
suggest that BM invariant mass spectra be investigated
extensively, where B is a baryon (e.g., $p$, $\Lambda$, $\Sigma$,
$\Xi$, $\Omega$, with charged particles and $\gamma\gamma$ in the
final state) and M is a meson (e.g., $\pi$, $\eta$, $\omega$,
$\phi$, $K$, $D$, $J/\psi$, $B$, $\Upsilon$, with charged
particles and $\gamma\gamma$ in the final state).

Fortunately, the detectors at LEP and B~Factories are suited for
these searches. For instance, due to high combinatoric background,
stringent selection criteria involving particle identification may
be required, testing for (in)consistency with the $p$, $K^{\pm}$,
and $\pi^{\pm}$ hypotheses. Measurements of dE/dx, for example, in
CLEOII~\cite{nim_320_66} and ALEPH~\cite{nim_A294_121} offer
excellent separation, although admittedly over a limited
soft-momentum spectrum for adequate $K^{\pm}-p$ separation. One of
the most notable improvements implemented in
CLEOIII~\cite{nim_384_61} with respect to CLEOII is the addition
of a Cherenkov detector, which is the primary particle
identification device for high momentum hadrons. The BaBar
detector provides high performance particle identification,
especially over $91\%$ of 4$\pi$ via analysis of Cherenkov
radiation in its DIRC detector~\cite{BaBarTech}. Belle possesses
particle identification capabilities utilizing dE/dx measurements
and Time-of-Flight scintillation counters for the low-momentum
regime and Cerenkov aerogel detectors for the high momentum
regime~\cite{nim_479_117}. The reconstruction of the various
species of mesons as described in from the extensive experience in
other analyses performed in the past by the experiments at LEP and
B~Factories.

\section{Conclusions}
\label{sec:conclusions}

Strong experimental evidence for a narrow exotic baryon resonance with strangeness
S = +1, $\Theta^+$, has created a great deal of excitement. The
observation of the exotic baryon $\Theta^+$ could have a number of
exciting implications. The idea the anti-decuplet may be
finally realized: The hint of $\Xi_{3 \over 2}$ observed by NA49
seems to support this interpretation. Furthermore, heavier
partners of $\Theta^+$ may exist in nature, $\Theta^0_c$ and
$\Theta^+_b$, where $\overline{s}$ is replaced by $\overline{c}$
and $\overline{b}$, respectively.

The exotic baryons $\Theta^+$ and $\Xi_{3 \over 2}$ may be
observed in $\gamma\gamma$ collisions at LEP and B~Factories
thanks to the relatively high content of strange quarks in the
photon. The observation of these exotic baryons in $e^+e^-$
annihilations may be possible provided that these are produced
copiously enough in hadronization of quarks produced in $Z$
decays. This hypothesis seems to be supported by the observation
of $\Theta^+$ in neutrino-induced charged-current interactions.

The observation of $\Theta^+$ (whether in $e^+e^-$ annihilations
or in $\gamma\gamma$ collisions) seems to  be a cornerstone for
observing further exotic baryon resonances at LEP and B~Factories.
If $\Theta^+$ is observed at LEP and B~Factories, a new rich
window of opportunity will open. Additionally, we suggest that a
more general search in baryon-meson invariant mass spectra be
performed.

\section{Acknowledgments}
\label{sec:acknowledgements}

We would like to thank A.~B\"{o}hrer, D.~Diakonov, Y.~Gao,
M.~Karliner, J.L.~Rosner, W.~Schweiger and T.T.~Wu for most useful
discussions. We would also like to thank K.~Cranmer, S.~Paganis,
W.~Wiedenmann and H.~Zobernig for their comments and suggestions
to the text. This work was supported in part by the United States
Department of Energy through Grant No. DE-FG0295-ER40896.

\bibliographystyle{zeusstylem}
\bibliography{penta}

\begin{mcbibliography}{10}

\bibitem{pan_66_1715}
V.V.~Barmin {\it et al.} (DIANA Collaboration),
\newblock  Phys. Atom. Nucl. 66 (2003) 1715\relax
\relax
\bibitem{prl_91_012002}
T.~Nakano {\it et al.} (LEPS Collaboration),
\newblock  Phys. Rev. Lett. {\bf 91}  (2003)~ 012002\relax
\relax
\bibitem{hep-ex_0307018}
S.~Stepanyan {\it et al.} (CLAS Collaboration),
\newblock  hep-ex/0307018\relax
\relax
\bibitem{hep-ex_0311046}
V.~Kubarovsky {\it et al.} (CLAS Collaboration),
\newblock  hep-ex/0311046\relax
\relax
\bibitem{hep-ex_0307083}
J.~Barth {\it et al.} (SAPHIR Collaboration),
\newblock  submitted to Phys. Lett. {\bf B}, hep-ex/0307083\relax
\relax
\bibitem{hep-ex_0309042}
A.E.~Asratyan, A.G.~Dolgolenko and M.A.~Kubantsev,
\newblock  submitted to Yad. Fiz. (Phys. At. Nucl.), hep-ex/0309042\relax
\relax
\bibitem{hep-ex_0312044}
A.~Airapetian {\it et al.} (HERMES Collaboration),
\newblock  hep-ex/0312044\relax
\relax
\bibitem{hep-ph_0311125}
A.R.~Dzierba {\it et al.},
\newblock  hep-ph/0311125\relax
\relax
\bibitem{Diak_Petrov_1984}
D.~Diakonov and V.~Petrov,
\newblock  preprint LNPI-967 (1984), published in Elementary Particles, Moscow,
  Energoatomizdat (1985) Vol. 2., 50 (in Russian)\relax
\relax
\bibitem{np_256_600}
M.~Chemtob,
\newblock  Nucl. Phys. {\bf B256}  (1984)~ 600\relax
\relax
\bibitem{Prasza_1987}
M.~Praszalowicz,
\newblock  Skyrmions and Anomalies, World Scientific (1987) 112\relax
\relax
\bibitem{Walliser_1992}
H.~Walliser,
\newblock  Baryon as Skyrme Soliton, World Scientific (1992) 247\relax
\relax
\bibitem{np_A548_649}
H.~Walliser,
\newblock  Nucl. Phys. {\bf A548}  (1984)~ 649\relax
\relax
\bibitem{Prasza_1987_talk}
M.~Praszalowicz,
\newblock  talk at workshop on Skyrmions and Anomalies, M.~Jezabek and
  M.~Praszalowicz editors, World Scientific 1987, 112\relax
\relax
\bibitem{hep-ph_0308114}
M.~Praszalowicz,
\newblock  hep-ph/0308114\relax
\relax
\bibitem{zp_A359_305}
D.~Diakonov, V.~Petrov and M.V.~Polyakov,
\newblock  Z. Phys. {\bf A359} (1997) 305\relax
\relax
\bibitem{Diakonov_1}
D.~Diakonov,
\newblock  private communication\relax
\relax
\bibitem{hep-ph_0310212}
D.~Diakonov and V.~Petrov,
\newblock  hep-ph/0310212\relax
\relax
\bibitem{hep-ph_0312325}
C.E.~Carlson, C.D.~Carone, H.J.~Kwee and V.~Nazaryan,
\newblock  hep-ph/0312325\relax
\relax
\bibitem{jhep_0311_70}
F.~Csikor, Z.~Fodor, S.D.Katz and T.G.~Kovacs,
\newblock  JHEP {\bf 0311} (2003) 070\relax
\relax
\bibitem{hep-lat_0310014}
S.~Sasaki,
\newblock  hep-lat/0310014\relax
\relax
\bibitem{hep-ex_0310014}
C.~Alt {\it et al.} (NA49 Collaboration),
\newblock  hep-ex/0310014\relax
\relax
\bibitem{hep-ph_0312369}
R.L.~Jaffe and F.~Wilczek,
\newblock  hep-ph/0312369\relax
\relax
\bibitem{hep-ph_0307341}
R.L.~Jaffe and F.~Wilczek,
\newblock  hep-ph/0307341\relax
\relax
\bibitem{hep-ph_0307343}
M.~Karliner and H.J.~Lipkin,
\newblock  hep-ph/0307343\relax
\relax
\bibitem{hep-ph_0308176}
K.~Cheung,
\newblock  hep-ph/0308176\relax
\relax
\bibitem{hep-ph_0312319}
A.K.~Leibovich {\it et al.},
\newblock  hep-ph/0312319\relax
\relax
\bibitem{hep-ex_0307088}
V.~Kubarovsky and S.~Stepanyan (for the CLAS Collaboration),
\newblock  hep-ex/0307088\relax
\relax
\bibitem{pr_50_5484}
M.~Artuso {\it et al.} (CLEO Collaboration),
\newblock  Phys. Rev. {\bf D50}  (1994)~ 5484\relax
\relax
\bibitem{pl_407_185}
H.~Hamasaki {\it et al.} (VENUS Collaboration),
\newblock  Phys. Lett. {\bf B407}  (1997)~ 185\relax
\relax
\bibitem{epj_28_45}
G.~Abbiendi {\it et al.} (OPAL Collaboration),
\newblock  Eur. Phys. J. {\bf C28}  (2003)~ 45\relax
\relax
\bibitem{pl_536_24}
M.~Achard {\it et al.} (L3 Collaboration),
\newblock  Phys. Lett. {\bf B536}  (2002)~ 24\relax
\relax
\bibitem{pl_571_11}
M.~Achard {\it et al.} (L3 Collaboration),
\newblock  Phys. Lett. {\bf B571}  (2003)~ 11\relax
\relax
\bibitem{Kuo_photon2003}
C.C.~Kuo {\it et al.} (for the Belle Collaboration),
\newblock  in proceedings of PHOTON 2003, Frascati, Italy, eds. F. Anulli {\it
  et al.}, to appear in Nucl. Phys. Proc. Suppl.\relax
\relax
\bibitem{pr_22_2157}
G.P.~Lepage and S.J.~Brodsky,
\newblock  Phys. Rev. {\bf D22}  (1980)~ 2157\relax
\relax
\bibitem{Lepage_Brodsky_1989}
G.P.~Lepage and S.J.~Brodsky,
\newblock  Exclusice Processes in Quantum Chromodynamics, in Perturbative
  Quantum Chromodynamics,
\newblock  World Scientific, Singapore 1989, 93\relax
\relax
\bibitem{epj_28_249}
C.F.~Berger and W.~Schweiger,
\newblock  Eur. Phys. J. {\bf C28}  (2003)~ 249\relax
\relax
\bibitem{hep-ph_0309095}
C.F.~Berger and W.~Schweiger,
\newblock  hep-ph/0309095\relax
\relax
\bibitem{Schweiger_1}
W.~Schweiger,
\newblock  private communication\relax
\relax
\bibitem{pr_40_2772}
H.~Aihara {\it et al.} (TPC/Two-Gamma Collaboration),
\newblock  Phys. Rev. {\bf D40}  (1989)~ 2772\relax
\relax
\bibitem{pr_37_28}
H.~Aihara {\it et al.} (TPC/Two-Gamma Collaboration),
\newblock  Phys. Rev. {\bf D37}  (1988)~ 28\relax
\relax
\bibitem{zp_C41_353}
W.~Braunschweig {\it et al.} (TASSO Collaboration),
\newblock  Z. Phys. {\bf C41} (1988) 353\relax
\relax
\bibitem{zp_C38_521}
C.~Berger {\it et al.} (PLUTO Collaboration),
\newblock  Z. Phys. {\bf C38} (1988) 521\relax
\relax
\bibitem{pl_198_255}
H.~Albrecht {\it et al.} (ARGUS Collaboration),
\newblock  Phys. Lett. {\bf B198}  (1987)~ 255\relax
\relax
\bibitem{zp_C32_11}
M.~Althoff {\it et al.} (TASSO Collaboration),
\newblock  Z. Phys. {\bf C32} (1986) 11\relax
\relax
\bibitem{hep-ph_0312269}
J.L.~Rosner,
\newblock  hep-ph/0312269\relax
\relax
\bibitem{prl_66_010001}
K.~Hagiwara {\it et al.} (The Particle Data Group),
\newblock  Phys. Rev. {\bf D66}  (2003)~ 010001\relax
\relax
\bibitem{hep-ph_0311355}
S.J~Brodsky,
\newblock  hep-ph/0311355\relax
\relax
\bibitem{nim_320_66}
Y.~Kuboda {\it et al.} (CLEO Collaboration),
\newblock  Nucl. Instrum. Methods {\bf A320}  (1992)~ 66\relax
\relax
\bibitem{nim_384_61}
S.E.~Kopp (for the CLEO Collaboration),
\newblock  Nucl. Instrum. Methods {\bf A384}  (1996)~ 61\relax
\relax
\bibitem{CLNS_01_1742}
R.A.~Briere {\it et al.} (CLEO-c Collaboration),
\newblock  CLNS-01-1742 (2001)\relax
\relax
\bibitem{BaBarTech}
BaBar Collaboration,
\newblock  BaBar Technical Design Report, SLAC-R-95-457 (1995)\relax
\relax
\bibitem{KEKB}
KEKB,
\newblock  B Factory Design Report, KEK Report 95-1 (1995), unpublished\relax
\relax
\bibitem{nim_479_117}
A.~Abashian {\it et al.} (Belle Collaboration),
\newblock  Nucl. Instrum. Methods {\bf A479}  (2002)~ 117\relax
\relax
\bibitem{jetp_14_1308}
V.N.~Gribov {et al.},
\newblock  Sov. Phys. JETP {\bf 14}  (1962)~ 1308\relax
\relax
\bibitem{pr_15_181}
V.M. Budnev, I.F. Ginzburg, G.V. Meledin and V.~G. Serbo,
\newblock  Phys. Rev. {\bf C15}  (1975)~ 181\relax
\relax
\bibitem{hep-ph_0311258}
J.J.~Dudek and F.E.~Close,
\newblock  Submitted to Elsevier Science, hep-ph/0311258\relax
\relax
\bibitem{prl_89_251802}
H.~Muramatsu {\it et al.} (CLEO Collaboration),
\newblock  Phys. Rev. Lett. {\bf 89}  (2002)~ 251802\relax
\relax
\bibitem{hep-ex_0308034}
K.~Abe {\it et al.} (Belle Collaboration),
\newblock  hep-ex/0308034\relax
\relax
\bibitem{prl_91_121801}
B.~Aubert {\it et al.} (BaBar Collaboration),
\newblock  Phys. Rev. Lett. {\bf 91}  (2003)~ 012002\relax
\relax
\bibitem{pr_65_091104}
B.~Aubert {\it et al.} (BaBar Collaboration),
\newblock  Phys. Rev. {\bf D65}  (2002)~ 091104\relax
\relax
\bibitem{zp_C69_393}
D.~Buskolic {\it et al.} (ALPEH Collaboration),
\newblock  Z. Phys. {\bf C69} (1996) 393\relax
\relax
\bibitem{zp_C73_601}
D.~Buskolic {\it et al.} (ALPEH Collaboration),
\newblock  Z. Phys. {\bf C73} (1997) 601\relax
\relax
\bibitem{pl_526_34}
A.~Heister {\it et al.} (ALPEH Collaboration),
\newblock  Phys. Lett. {\bf B526}  (2002)~ 34\relax
\relax
\bibitem{nim_A294_121}
ALEPH Collaboration,
\newblock  Nucl. Instrum. Methods {\bf A294}  (1990)~ 121\relax
\relax
\end{mcbibliography}

\end{document}